\documentclass[twocolumn]{jpsj2}
\title{
Quasiclassical Theory for 
Thermodynamic Properties of the Two-Band Superconductor $\mbox{MgB}_{2}$
under Magnetic Fields} 

\author{Masatoshi \textsc{Mugikura}\thanks{E-mail: mugikura@cmpt.phys.tohoku.ac.jp}, Hiroaki \textsc{Kusunose} and Yoshio \textsc{Kuramoto}}

\inst{Department of Physics, Tohoku University, Sendai 980-8578}
\abst{
Consistent account is provided of strongly anisotropic properties in thermodynamics of $\mbox{MgB}_{2}$ under magnetic field $H$ using approximate analytic solution in the quasiclassical formalism.
With the strength of the pairing interactions and the Fermi velocities of two bands as fitting parameters, it is possible to 
reproduce in wide regions of the phase diagram the $H$- and temperature 
dependences of relevant quantities such as
the zero-energy density of states (ZEDOS),  two energy gaps,
the specific heat and the $H_{c2}$ anisotropy.
It is demonstrated that the velocity ratio of two bands affects considerably the low-field behavior of the ZEDOS and 
energy gaps. 
The ratio also affects the high-temperature behavior of the $H_{c2}$ anisotropy.
The anisotropy of the coherence length in the $\sigma$ band 
depends weakly on $H$ and $T$, 
whereas the coherence length in the $\pi$ band remains isotropic 
even in the presence of the inter-band coupling.
}

\kword{$\mbox{MgB}_{2}$, multiband superconductivity, specific heat, upper critical field, quasiclassical theory}

\begin{document}
\maketitle
\section{Introduction}

Superconductivity in $\mbox{MgB}_{2}$\cite{nagamatsu} has attracted much interest not only for its high transition temperature $T_{c}=39$K but for 
the unique character coming from 
distinct Fermi surfaces in $\sigma$ and $\pi$ bands.
Owing to 
the presence of multiple bands, the superconducting 
state has two different energy gaps, although the pairing is triggered by the electron-phonon coupling in the $\sigma$ band.
The 
two-band nature of superconductivity has been supported by various experiments such as point contact spectroscopy\cite{shabo,gonneli},
 specific heat\cite{wang,bouquet,bouquetsh}, and the angle-resolved photoemission spectroscopy (ARPES)\cite{souma}. 
In addition to providing two different gaps, the $\sigma$ and $\pi$ bands have different dimensionality, i.e., the former is quasi two-dimensional and the latter is rather three-dimensional.
The interplay of the two gaps with different size and anisotropy 
should lead to characterisitic behaviors, particularly in thermodynamic quantities in a mixed state.
Indeed, the strong temperature dependence of the anisotropy of the upper critical field, $\Gamma(T)=H_{c2}^{(ab)}(T)/H_{c2}^{(c)}(T)$, has been observed\cite{angst,eltsev,lyard} and has been discussed theoretically as a consequence of such interplay\cite{miranovic,dahm,gurevich,golubov,zhitomirski}.

There have been a number of theoretical investigations on various 
thermodynamic properties in the mixed state, such as 
the zero-temperature specific-heat coefficient $\gamma (H)$ \cite{nakai,tewordt} and the field dependence of the energy gaps\cite{graser,koshelev,ichioka}.
However, validity of each theoretical tool used in these studies is limited to either $H\sim H_{c2}$ or $T \sim 0$.
Hence consistent account of the whole $H$-$T$ phase diagram within the single framework is highly desired.

Recently, one of the present authors has developed a theoretical tool to investigate thermodynamic properties\cite{kusu} using the approximate analytic solution in the quasiclassical formalism\cite{pesch}.
The approximate solution was originally proposed by Brandt, Pesch and Tewordt (BPT) in the Gor'kov formalism\cite{bpt},
and is most suitable to strongly type-II superconductors near $H_{c2}$. 
However, a comparison with reliable numerical calculations indicates that the present approximation is applicable semi-quantitatively to the wide region of the $H$-$T$ phase diagram\cite{kusu}.
The purpose of this paper is to give a consistent account of thermodynamic properties based on the approximate analytic solution, especially of anisotropic behaviors in the mixed phase.

The paper is organized as follows.
In the next section, we briefly introduce the method of  approximate analytic solution.
In $\S$ 3, we first determine the plausible strength of pairing interactions at $H=0$, and discuss the ratio of the Fermi velocities of the two bands focusing on the temperature dependence of the anisotropy of $H_{c2}$.
Then, using the parameters obtained, we compare $H$ and $T$ dependence of the specific heat with experimental results.
It is shown that the low-$H$ behavior of the zero-energy density of states (ZEDOS) and the gaps is sensitive to the ratio of the velocities of two bands.
The $H$ and $T$ dependence of the anisotropy of the coherence length is also discussed.
The last section summarizes the paper.

\section{The BPT approximation}
To discuss thermodynamics under magnetic fields we have to solve the Eilenberger equation, the Maxwell equation and the gap equation self-consistently.
It is tedious to solve these equations as they stand.
However, if we approximate the spatial dependence of the internal field and the diagonal part of the quasiclassical Green function by their spatial average, ${\bf B}$ and $g$, and 
assume the Abrikosov lattice for the spatial variation of the gap function, we obtain the analytic solution of 
the quasiclassical Green function. 
Using $\hbar=k_{B}=c=1$ hereafter, the result is given by \cite{kusu} 
\begin{equation}
g_{\ell}=\Bigl[ 1+\frac{\sqrt{\pi}}{i}\Bigl(\frac{u_{n\ell}\Delta_\ell}{\omega_n}\Bigr)^{2}  W^{'}(iu_{n\ell}) \Bigr]^{-1/2},
\label{g_l}
\end{equation}
where $\ell$ is the band index (either $\sigma$ or $\pi$), 
$\omega_n$  is the fermionic Matsubara frequency,
$W(z)=e^{-z^2}{\rm erfc}(-iz)$ is the Faddeeva function, 
and $\Delta_{\ell}$ is the spatial average of the energy gap in the $\ell$-band.
We have assumed the isotropic $s$-wave gap for both bands.
In eq.(\ref{g_l}) we have introduced the quantity
\begin{equation}
u_{n\ell}=\frac{2\omega_n}{\tilde{v}_{\ell \perp}(\hat{\mib k})\sqrt{2|e|B}} ,
\end{equation}
where $\hat{{\mib k}}={\mib k}_{F}/|{\mib k}_{F}|$ is the normalized Fermi wave vector,  
and $\tilde{v}_{\ell \perp}(\hat{\mib k})$ is related to the Fermi velocity perpendicular to the applied field ${\mib H}$.
We shall give explicit expression of $\tilde{v}_{\ell \perp}(\hat{\mib k})$ later.

The Maxwell equation and the gap equation follow from
the stationary condition of the free energy with respect to $B$ and $\Delta_\ell$, respectively. 
In the BPT approximation the free energy measured from that of the normal state in the clean limit\cite{kusu} is given by
\begin{multline}
\label{freeenergy}
\Omega_{SN}=
\sum_{\ell,\ell'}(\hat{V}^{-1})_{\ell\ell'}\Delta_{\ell}\Delta_{\ell'}+\sum_{\ell}N_{0\ell}\Bigl[ \Delta_{\ell}^{2}\ln \Bigl(\frac{T}{T_{c}}\Bigr)\\
-\Delta_{\ell}^{2}\ln \Bigl( \frac{2 e^{\gamma} \omega_{c}}{\pi T_{c}}\Bigr)
+2\pi T\sum_{n=0}^{\infty}\Bigl( \frac{\Delta_{\ell}^{2}}{\omega_{n}}-\langle I_{\ell} \rangle_{\ell} \Bigr) \Bigr],
\end{multline}
where 
\begin{equation}
I_{\ell}=\frac{2\sqrt{\pi}g_{\ell}}{1+g_{\ell}}\frac{u_{n\ell}}{\omega_n} \Delta_{\ell}^{2}
  W(iu_{n\ell}),
\end{equation}
and $\langle \cdots \rangle_{\ell}$ denotes the average over the Fermi surface. 
In eq.(\ref{freeenergy}),
$N_{0\ell}$ is the DOS in the normal state, and $\gamma\simeq 0.5772$ is the Euler's constant. 
The cut-off frequency $\omega_{c}=60\pi T_c$ is used in the present paper.
We have assumed $B\sim H$ since $\mbox{MgB}_2$ is the strongly type-II superconductor with $\kappa\sim 26$\cite{kappa}.

For simplicity, we use the cylindrical (spherical) Fermi surface for the two-dimensional (three-dimensional) $\sigma$ ($\pi$) band. 
Then the Fermi velocities are given by
\begin{eqnarray}
\label{sigma}
&&{\mib v}_{\sigma}(\hat{{\mib k}})=\bigl( v_{\sigma ,ab} \cos \phi,v_{\sigma ,ab} \sin \phi,v_{\sigma ,c}\mbox{sgn}(k_{c})\bigr), \\
&&{\mib v}_{\pi}(\hat{{\mib k}})=v_{\pi}(\sin \theta \cos \phi, \sin \theta \sin \phi, \cos \theta),
\end{eqnarray}
where $\phi$ ($\theta$) is the azimuthal (polar) angle of $\hat{{\mib k}}$.
Here the quasi two-dimensionality in $\sigma$ band is taken into account 
simply by $v_{\sigma,c} \mbox{sgn}(k_{c})$ with $v_{\sigma,c}\ll v_{\sigma,ab}$.
We introduce $\tilde{v}_{\ell\perp}(\hat{\mib k})$ by
\begin{equation}
\tilde{v}_{\ell\perp}(\hat{\mib k})=\sqrt{\chi_\ell^{-1/2}v_{\ell x}^2+\chi_\ell^{1/2}v_{\ell y}^2},
\end{equation}
for ${\mib H}\parallel{\mib c}$, and
\begin{equation}
\tilde{v}_{\ell\perp}(\hat{\mib k})=\sqrt{\chi_\ell^{-1/2}v_{\ell z}^2+\chi_\ell^{1/2}v_{\ell y}^2},
\end{equation}
for ${\mib H}\perp{\mib c}$.
Here $\chi_\ell$ characterizes the anisotropy of the coherence length perpendicular to the field.
Thus, $\chi_\ell=1$ for ${\mib H}\parallel{\mib c}$.
For ${\mib H}\perp{\mib c}$ there is strong anisotropy in $\sigma$ band due to the quasi two-dimensionality.
In this case, we determine $\chi_\ell=(\xi_{\ell z}/\xi_{\ell y})^2$ variationally by minimizing the free energy\cite{dahm,udagawa}.
In the case of the Ginzburg-Landau theory, $\chi_{\ell {\rm GL}}=\langle v_{\ell z}^{2}\rangle/\langle v_{\ell y}^{2} \rangle$.

Due to the coupling between $\sigma$ and $\pi$ bands, it may 
be expected that $\chi_\pi$ differs from unity even if $\pi$ band is isotropic.
It turns out, however, that the variational solution has no anisotropy in $\chi_\pi$ within the numerical accuracy, whereas $\chi_\sigma$ has a strong anisotropy. 
This result is consistent with that obtained by Ichioka \textit{et al.}, in which the local DOS(LDOS) of the $\pi$ band is nearly isotropic while that of the $\sigma$ band has strong anisotropy\cite{ichioka}.
On the experimental side,
almost isotropic vortex core in the $\pi$ band is observed by scanning tunneling spectroscopy(STS) at low fields, even in the case of ${\mib H}\perp {\mib c}$\cite{eskildsen}.
On the other hand, the use of the Abrikosov lattice with different $\chi_\ell$'s in two bands results in unphysical situation since the structure of penetrating flux differs in two bands.
However, after spatial average over a larger scale,  this deficiency of the solution may not be serious as far as we are concerned with homogeneous properties. 
Since only spatially averaged quantities have a meaning in the BPT approximation, we use
the variational solution, $\chi_\pi=1$, throughout this paper.

\section{Anisotropy of thermodynamic quantities}

\begin{figure}[t]
\includegraphics[width=8cm]{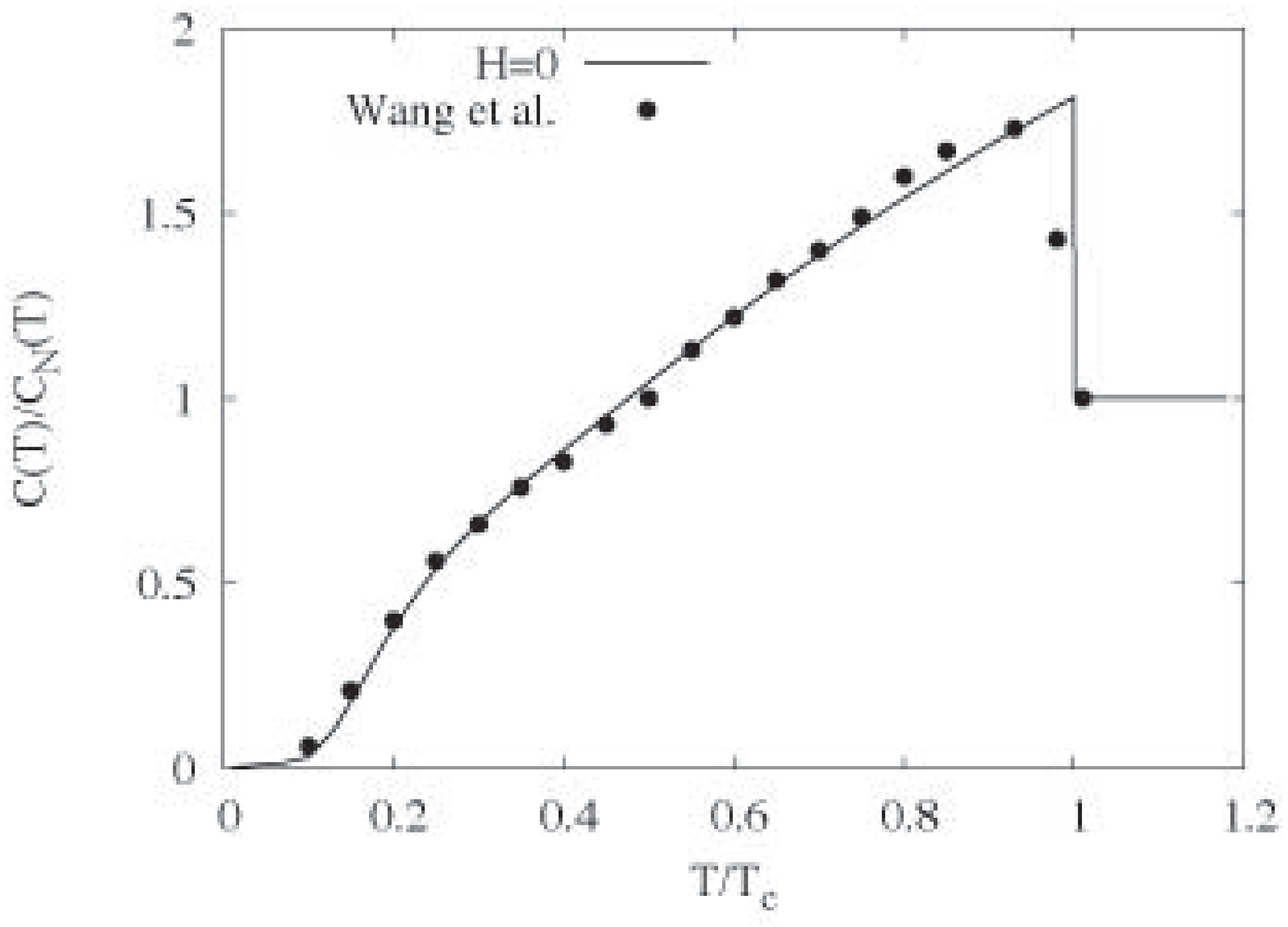}
\caption{The temperature dependence of the specific heat at $H=0$. The solid circle represents the experimental data 
of Wang \textit{et al}\cite{wang}.}
\label{specificheat}
\end{figure}

Let us first determine the strength of the pairing interaction in the matrix,
\begin{equation}
\hat{V}=\begin{pmatrix} V_{\sigma} & V \\ V & V_{\pi} \end{pmatrix},
\end{equation}
by fitting the specific heat data at $H=0$.
Note that the present approximation recovers the ordinary two-band BCS theory\cite{suhl,moskalenko} at $H=0$.
We use $N_{0\sigma}/N_{0}=0.45$ and $N_{0\pi}/N_{0}=0.55$ ($N_{0}$ is the total DOS) from the specific heat measurement
\cite{bouquetsh}. These values are in good agreement with the band structure calculation.\cite{Dos}
Figure~\ref{specificheat} shows the best fitting of the $T$ dependence of the specific heat at $H=0$. 
The coupling constants are determined to be $N_{\sigma}V_{\sigma}=0.30$, $V_{\pi}/V_{\sigma}=0.14$ and $V/V_{\sigma}=0.34$.
The solid circle is taken from the experimental data by Wang \textit{et al.}\cite{wang}
We obtain the gap magnitude at $T=0$ as $\Delta_{\sigma 0}=6.8$ meV
and $\Delta_{\pi 0}=2.4$ meV, which are in good agreement with those determined by tunneling spectroscopy \cite{shabo,gonneli}.
Our coupling constants are slightly different from those determined by the band structure calculations\cite{liu,golubovband} due to the simplification of the Fermi surfaces in the present study.
In this work we try to reproduce of overall $T$ dependence of the gap magnitude, i.e., specific heat,  rather than determining the coupling constants uniquely.
In fact, different choice of coupling constants may give similar behaviors in $C(T)$\cite{zhitomirski,watanabe}.

\begin{figure}[t]
\includegraphics[width=8cm]{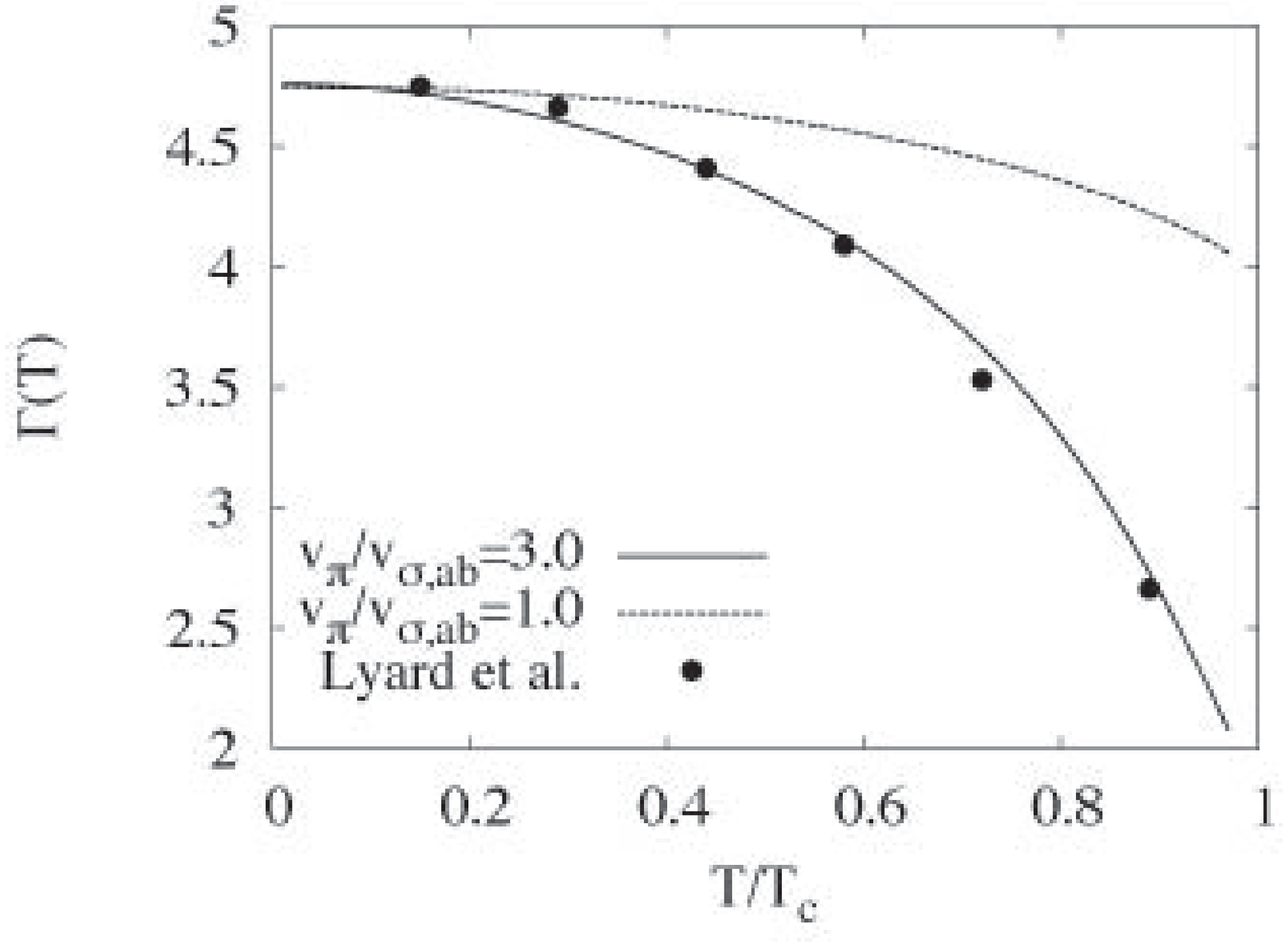}
\caption{The ratio 
$\Gamma$ of the upper critical fields $H_{c2}(T)$ along the $c$-axis and the $ab$-plane against temperature. The solid circles are the experimental data
by Lyard \textit{et al}.\cite{lyard}}
\label{gamma}
\end{figure}
Next, we determine the Fermi velocities of two bands by fitting the anisotropy of the upper critical fields, $\Gamma(T)$.
The solid line in Fig.~\ref{gamma} represents our result for $v_{\sigma,c}/v_{\sigma,ab}=0.13$ and $v_{\pi}/v_{\sigma,ab}=3.0$, while the solid circles are experimental data taken from Lyard \textit{et al.}\cite{lyard}
The result for $v_{\sigma,c}/v_{\sigma,ab}=0.13$, $v_{\pi}/v_{\sigma,ab}=1.0$ is also shown in Fig.~\ref{gamma} (dashed line) for comparison.
The slight modification of the 
velocity ratio gives a considerable change in $\Gamma(T)$ especially at high temperatures.
The present result for $\Gamma(T)$ is consistent with previous theoretical works\cite{dahm,miranovic,gurevich,golubov,zhitomirski}.
It is emphasized that the strong temperature dependence of $\Gamma(T)$ originates from the interplay of the two gaps since the $T$ dependence becomes much weaker in the absence of the interband coupling, $V=0$.
Note that $\Gamma(T)$ is independent of $T$ in the single-band anisotropic GL theory.
The values of  $v_{\sigma,c}/v_{\sigma,ab}=0.13$ and $v_{\pi}/v_{\sigma,ab}=3.0$ are in good agreement with those obtained from the band structure calculation\cite{Fermi,Dos}.

\begin{figure*}
\begin{center}
\includegraphics[width=15cm]{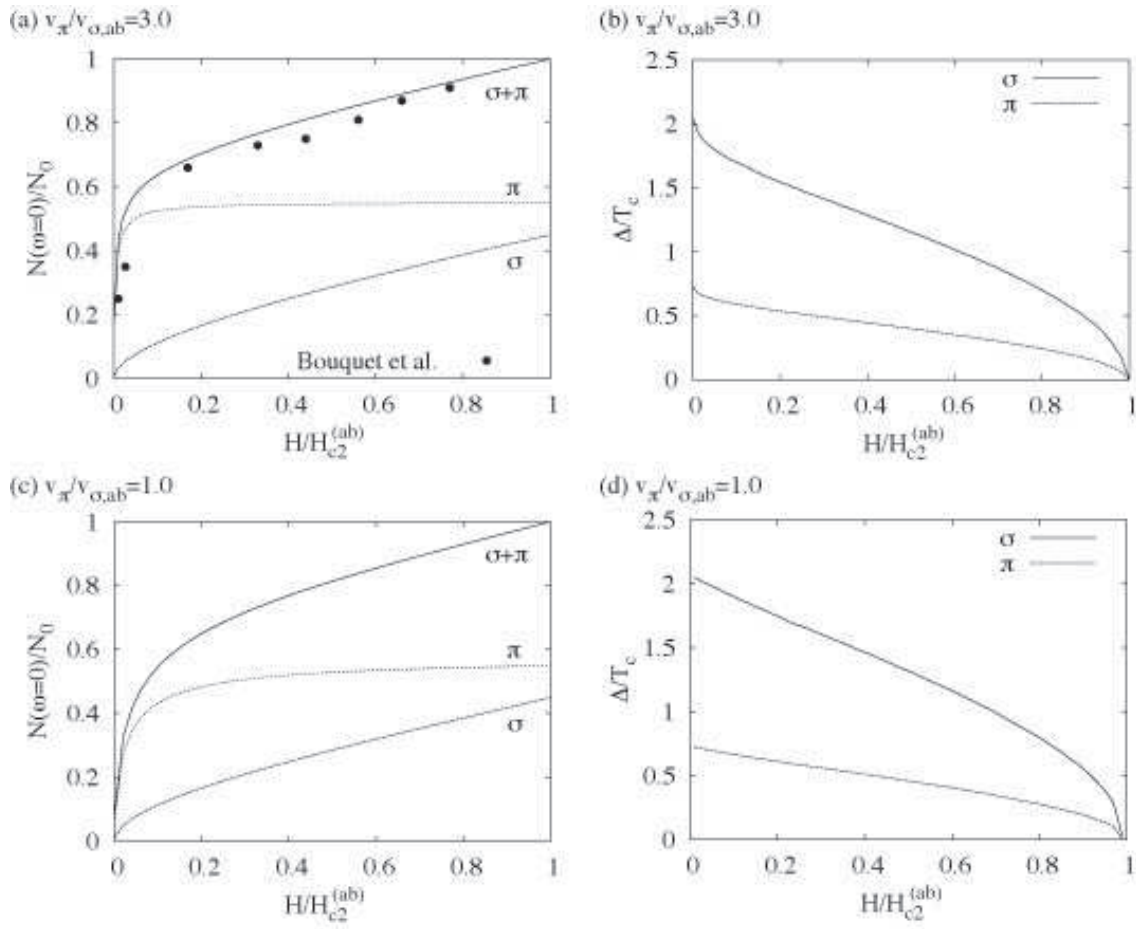}
\caption{The in-plane field dependence of (a) the ZEDOS, and (b) the two gaps at $T/T_{c}=0.05$.
The solid circle represents the experimental data taken from Bouquet \textit{et al}.\cite{bouquetsh}
For comparison the results for $v_{\pi}/v_{\sigma,ab}=1.0$ are shown in (c) for ZEDOS and (d) for the gaps.} 
\label{gapdos}
\end{center}
\end{figure*}

With these preliminaries, we discuss the field dependence using the coupling constants and the Fermi velocities determined 
above.
Figure~\ref{gapdos} shows the in-plane field dependence of two gaps and the ZEDOS at $T/T_{c}=0.05$.
The solid circle denotes the experimental data taken from Bouquet \textit{et al.}\cite{bouquetsh}
Owing to the smallness of $\Delta_{\pi}$ as compared with $\Delta_{\sigma}$, the coherence length $\xi_\pi$ in $\pi$ band is much larger than $\xi_\sigma$. Indeed, we obtain $\xi_{\pi}/\xi_{\sigma,ab} \sim 7$ and $\xi_{\pi}/\xi_{\sigma,c} \sim 60$ for $v_{\pi}/v_{\sigma,ab}=3.0$ using the BCS formula.
On the other hand, the upper critical field is determined mostly by the pairing in the $\sigma$ band.
Thus, the rapid rise of the ZEDOS should result as shown in Fig.~\ref{gapdos}(a).
The contribution of the $\pi$ band to the ZEDOS saturates already at $H/H_{c2}^{(ab)}\sim 0.05$.
Note that $\Delta_{\pi}$ 
remains nonzero up to $H_{c2}$ because of the interband coupling as shown in Fig.~\ref{gapdos}(b).

The results in the case of $v_{\pi}/v_{\sigma,ab}=1.0$ are shown in Fig.\ref{gapdos}(c) and (d) for comparison.
The change of the 
velocity ratio only affects the low-field behaviors of the ZEDOS and of the gaps.
On the contrary, $\Gamma(T)$ of the $H_{c2}$ ratio is affected much in higher temperatures as shown in Fig.~\ref{gamma}.
Experimentally Gonneli \textit{et al.}\cite{gonnelimag} report the rapid suppression of the gaps, while Bugolavsky \textit{et al.}\cite{bugolavsky} report no considerable suppression at low fields.
There seems sample dependence in $\xi_{\pi}/\xi_{\sigma}$ presumably due to the impurities.
There are a number of investigations on the field dependence of the ZEDOS and the gaps both theoretically\cite{koshelev,graser,ichioka} and experimentally\cite{gonnelimag,bugolavsky}.
All of these show similar behaviors except in low-field region.

\begin{figure}[h]
\includegraphics[width=8cm]{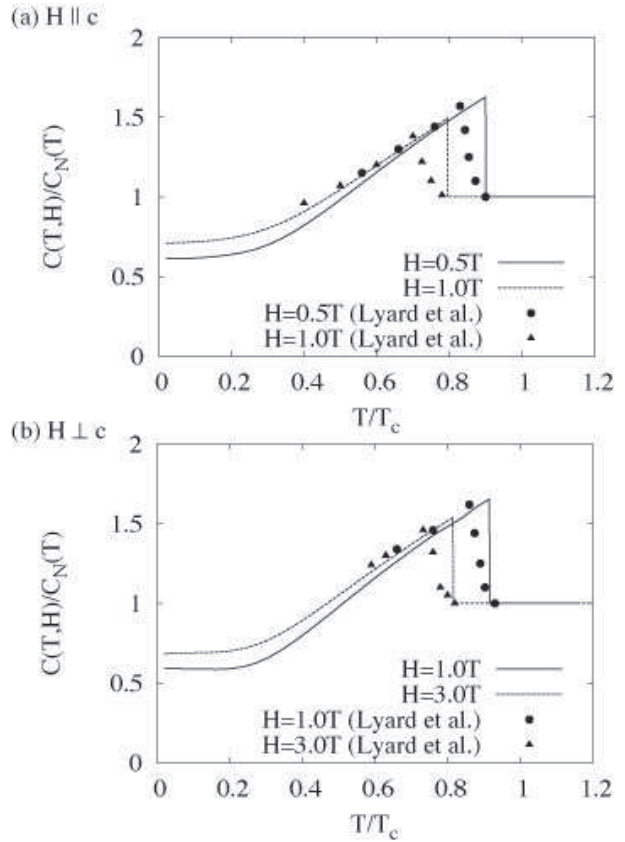}
\caption{The $T$ dependence of the specific heat under magnetic fields (a) for ${\mib H}\parallel{\mib c}$ and (b) for ${\mib H}\perp{\mib c}$.
The experimental data represented by symbols are taken from Lyard \textit{et al.}\cite{lyard} The experimental value of $H_{c2}$ at $T=0$ is $H_{c2}^{(ab)}=17$T and $H_{c2}^{(c)}=3.5$T.} 
\label{specificmag}
\end{figure}

Once we determine the necessary parameters, we calculate various quantities under magnetic fields without any free parameters.
The $T$ dependences of the specific heat under magnetic fields with $v_{\pi}/v_{\sigma,ab}=3.0$ are shown in Fig.~\ref{specificmag} for (a) ${\mib H}\parallel {\mib c}$ and (b) for ${\mib H}\perp {\mib c}$.
For both field directions, the overall behavior of $C(T,H)$ is consistent with the experimental results indicated in the figures.
In particular, the observed specific heat jump at $T_{c}(H)$ is reproduced very well for both field directions.

\begin{figure}
\includegraphics[width=8cm]{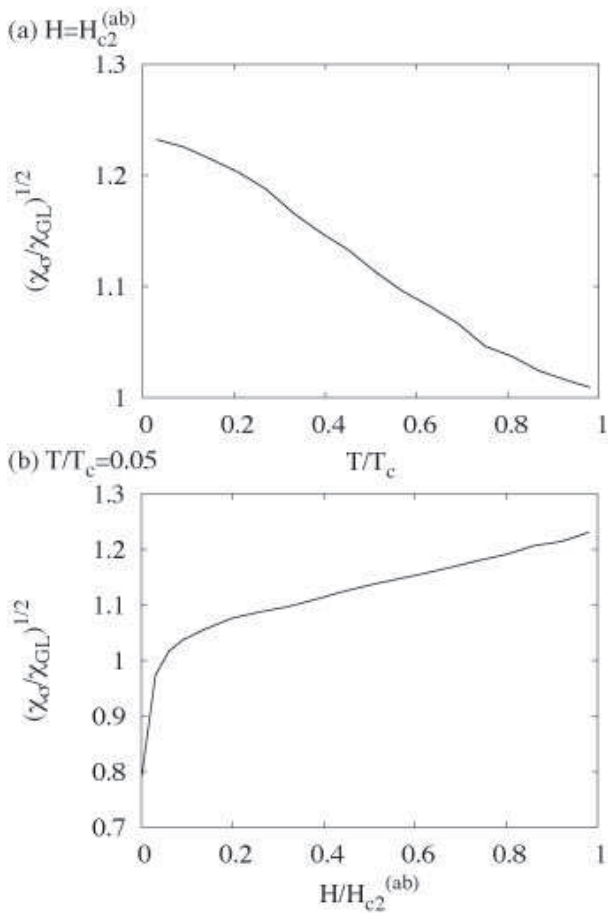}
\caption{(a) The temperature dependence of $\chi_{\sigma}^{1/2}$ at $H=H_{c2}^{(ab)}$.
(b) The field dependence of $\chi_{\sigma}^{1/2}$ at $T/T_{c}=0.05$.
The anisotropy in the GL theory is given by $\chi_{GL}^{1/2}$.} 
\label{chi}
\end{figure}

Finally, we discuss the $H$ and $T$ dependence of $\chi_\sigma$ for in-plane fields.
The $T$ dependence of $\chi_\sigma$ at $H=H_{c2}^{(ab)}$ is shown in Fig.~\ref{chi}(a), where $\chi_{GL}^{1/2}=\sqrt{2} v_{\sigma,c}/v_{\sigma,ab}\simeq 0.18$.
At low temperatures the vortex cores elongate 
along $c$-axis slightly more 
as compared with the GL theory.
In the limit of $T\to T_c$, the anisotropy recovers the GL value.
The vortex cores also elongate monotonously along $c$-axis with increase of the field as shown in Fig.~\ref{chi}(b).
It should be noted that the $T$ and $H$ dependence of $\chi_\sigma$ does not come from the effect of two-band superconductivity, but comes purely from the quasi two-dimensionality of the $\sigma$ band.
This conclusion follows since there is no essential difference even in the absence of the inter-band interaction, $V=0$.

\section{Summary}
We have investigated the thermodynamic properties and its anisotropy of the two-band superconductivity in $\mbox{MgB}_{2}$ on the basis of the approximate analytic solution in the quasiclassical formalism.
Using the strength of the pairing interactions and the Fermi velocities of two bands as fitting parameters,
we have reproduced the $T$- and $H$- dependence of 
the ZEDOS, the two gaps, the specific heat, and the anisotropy of the upper critical field in wide regions of the phase diagram.

The change of the velocity ratio affects considerably the low-field behaviors of the ZEDOS and the energy gaps.
At the same time the velocity ratio determines  the high-temperature behavior of the $H_{c2}$ anisotropy $\Gamma(T)$.
Thus, our theory predicts a strong correlation between the low-field behaviors of the ZEDOS and the high-temperature behavior of $\Gamma(T)$. 
We have shown that the anisotropy of the coherence length in the $\sigma$ band has $H$ and $T$ dependence different from that in the GL theory.
The isotropic nature of the coherence length in the $\pi$ band remains even if the inter-band coupling is present.


\begin{thebibliography}{99}
\bibitem{nagamatsu}
J. Nagamatsu, N. Nakagawa, T. Muranaka, Y. Zenitani and J. Akimitsu: Nature $\textbf{410}$ (2001) 63.
\bibitem{shabo}
P. Szab\'{o}, P. Samuely, J. Ka\v{c}mar\v{c}\'{i}k, T. Klein, J. Marcus, D. Fruchart, S. Miraglia, C. Marcenat and A. G. M. Jansen:
Phys. Rev. Lett. $\textbf{87}$ (2001) 137005. 
\bibitem{gonneli}
R. S. Gonnelli, D. Daghero, G. A. Ummarino, V. A. Stepanov, J. Jun, S. M. Kazakov and J. Karpinski: Phys. Rev. Lett. $\textbf{89}$ (2002) 247004. 
\bibitem{wang}
Y. Wang, T. Plackowski and A. Junod: Physica C $\textbf{355}$ (2001) 179. 
\bibitem{bouquet}
F. Bouquet, R. A. Fisher, N. E. Phillips, D. G. Hinks and J. D. Jorgensen: Phys. Rev. Lett. $\textbf{87}$ (2001) 047001.
\bibitem{bouquetsh}
F. Bouquet, Y. Wang, I. Sheikin, T. Plackowski, A. Junod, S. Lee and S. Tajima: Phys. Rev. Lett. $\textbf{89}$ (2002) 257001. 
 \bibitem{souma}
S. Souma, Y. Machida, T. Sato, T. Takahashi, H. Matsui, S. -C. Wang, H. Ding, A. Kaminski, J. C. Campuzano, S. Sasaki and K. Kadowaki:
Nature $\textbf{423}$ (2003) 65. 
\bibitem{angst}
M. Angst, R. Puzniak, A. Wisniewski, J. Jun, S. M. Kazakov, J. Karpinski, J. Roos and H. Keller: Phys. Rev. Lett. $\textbf{88}$ (2002) 167004. 
\bibitem{eltsev}
Yu. Eltsev, S. Lee, K. Nakao, N. Chikumoto, S. Tajima, N. Koshizuka and M. Murakami: Phys. Rev. B $\textbf{65}$ (2002) 140501(R). 
\bibitem{lyard}
L. Lyard, P. Samuely, P. Szab\'{o}, T. Klein, C. Marcenat, L. Paulius, K. H. P. Kim, C. U. Jung, H. -S. Lee,
B. Kang, S. Choi, S. -I.Lee, J. Marcus, S. Blanchard, A.G. M. Jansen, U. Welp, G. Karapetrov and W. K. Kwok: Phys. Rev. B $\textbf{66}$ (2002) 180502(R).
\bibitem{dahm}
T. Dahm and N. Schopohl: Phys. Rev. Lett. $\textbf{91}$ (2003) 017001. 
\bibitem{miranovic}
P. Miranovi\'{c}, K. Machida and V. G. Kogan: J. Phys. Soc. Jpn. $\textbf{72}$ (2003) 221. 
\bibitem{gurevich}
A. Gurevich: Phys. Rev. B $\textbf{67}$ (2003) 184515. 
\bibitem{golubov}
A. A. Golubov and A. E. Koshelev: Phys. Rev. B $\textbf{68}$ (2003) 104503. 
\bibitem{zhitomirski}
M. E. Zhitomirsky and V. -H. Dao: Phys. Rev. B $\textbf{69}$ (2004) 054508. 
\bibitem{nakai}
N. Nakai, M. Ichioka and K. Machida: J. Phys. Soc. Jpn. $\textbf{71}$ (2002) 23.
\bibitem{tewordt}
L. Tewordt and D. Fay: Phys. Rev. B $\textbf{67}$ (2003) 134524.
\bibitem{graser}
S. Graser, T. Dahm and N. Schopohl: Phys. Rev. B $\textbf{69}$ (2004) 014511.
\bibitem{koshelev}
A. E. Koshelev and A. A. Golubov: Phys. Rev. Lett. $\textbf{90}$ (2003) 177002.
\bibitem{ichioka}
M. Ichioka, K. Machida, N. Nakai and P. Miranovi\'{c}: Phys. Rev. B $\textbf{70}$ (2004) 144508.
\bibitem{eskildsen}
M.R. Eskildsen, N. Jenkins, G. Levy, M. Kulger, \O. Fischer, J. Jun, S.M. Kazakov and J. Karpinski: Phys. Rev. B $\textbf{68}$ (2003) 100508(R).
\bibitem{kusu}
H. Kusunose: Phys. Rev. B $\textbf{70}$ (2004) 054509, and references therein
\bibitem{bpt}
U. Brandt, W. Pesch and L. Tewordt: Z. Phys. $\textbf{201}$ (1967) 209. 
\bibitem{pesch}
W. Pesch: Z. Phys. B $\textbf{21}$ (1975) 263. 
\bibitem{kappa}
D.K. Finnemore, J.E. Ostenson, S.L. Bud'ko, G. Lapertot and P.C. Canfield: Phys. Rev. Lett. $\textbf{86}$ (2001) 2420. 
\bibitem{udagawa}
M. Udagawa, Y. Yanase and M. Ogata: Phys. Rev. B $\textbf{71}$ (2005) 024511.
\bibitem{Fermi}
A. Brinkman, A.A. Golubov, H. Rogalla, O.V. Dolgov, J. Kortus, Y. Kong, O. Jepsen and O.K. Andersen
: Phys. Rev. B $\textbf{65}$ (2002) 180517(R). 
\bibitem{Dos}
Y.Kong, O.V. Dolgov, O. Jepsen and O.K. Andersen: Phys. Rev. B $\textbf{64}$ (2001) 020501(R).
\bibitem{watanabe}
K. Watanabe and T. Kita: J. Phys. Soc. Jpn.$\textbf{73}$ (2004) 2239.
 \bibitem{suhl}
H. Suhl, B. T. Mattias and L. R. Walker: Phys. Rev. Lett. $\textbf{3}$ (1959) 552.
\bibitem{moskalenko}
V. A. Moskalenko: Fiz. Met. Metalloved. $\textbf{4}$ (1959)  503. [Phys. Met. Metallogr. (USSR) $\textbf{8}$ (1959) 25.]
\bibitem{liu}
A.Y Liu, I.I. Mazin and J. Kortus: Phys. Rev. Lett $\textbf{87}$ (2001) 087005.
\bibitem{golubovband}
A.A. Golubov, J. Kortus, O.V. Dolgov, O. Jepsen, Y. Kong, O.K. Andersen, B. J. Gibson, K. Ahn and R.K. Kremer: J. Phys. Condens. Matter $\textbf{14}$ (2002) 1353. 
\bibitem{gonnelimag}
R. S. Gonnelli, D. Daghero, A. Calzolari, G. A. Ummarino, Valeria Dellarocca, V. A. Stepanov, J. Jun, S. M. Kazakov and J. Karpinski:
Phys. Rev. B $\textbf{69}$ (2004) 100504(R).
\bibitem{bugolavsky}
Y. Bugolavsky, Y. Miyoshi, G. K. Perkins, A. D. Caplin, L. F. Cohen, A. V. Pogrebnyakov and X. X. Xi:
 Phys. Rev. B $\textbf{69}$ (2004) 132508.
\end{thebibliography}
\end{document}